\documentclass[twocolumn,english]{revtex4-1}
\usepackage[L7x,T1]{fontenc}
\usepackage[latin9]{inputenc}
\usepackage{verbatim}
\usepackage{amsmath}
\usepackage{amssymb}
\usepackage{graphicx}

\makeatletter
\relax

\makeatother

\usepackage{babel}
\begin{document}
\title{
\global\long\def\cm{\text{cm}^{-1}}%
\global\long\def\ips{\text{ps}^{-1}}%
 Temperature Controlled Open Quantum System Dynamics using Time-dependent
Variational Method}
\author{Mantas Jaku\v{c}ionis\textsuperscript{1}, Darius Abramavi\v{c}ius\textsuperscript{1}}
\affiliation{\textsuperscript{1}Institute of Chemical Physics, Vilnius University,
Sauletekio Ave. 9-III, LT-10222 Vilnius, Lithuania}
\begin{abstract}
Dirac-Frenkel variational method with Davydov $\text{D}_{2}$ trial
wavefunction is extended by introducing a thermalization algorithm
and applied to simulate dynamics of a general open quantum system.
The algorithm allows to control temperature variations of a harmonic
finite size bath, when in contact with the quantum system. Thermalization
of the bath vibrational modes is realised via stochastic scatterings,
implemented as a discrete-time Bernoulli process with Poisson statistics.
It controls bath temperature by steering vibrational modes' evolution
towards their canonical thermal equilibrium. Numerical analysis of
the exciton relaxation dynamics in a small molecular cluster reveals
that thermalization additionally provides significant calculation
speed up due to reduced number of vibrational modes needed to obtain
the convergence.
\end{abstract}
\maketitle

\section{Introduction}

Obtaining dynamics of open quantum systems, \textit{i.e.}, quantum
systems that are identified as separate from its environment, yet
remain in thermal contact with it, is one of the most general quantum
mechanical problems. Its applicability range from excited state relaxation
in optical response \citep{Dorfman2016,Mukamel1995}, energy transport
in molecular aggregates \citep{Jakucionis2018a,Chorosajev2017a,Schroter2015,Chorosajev2014a,Valkunasa,May2011},
photosynthetic complexes \citep{RevModPhys.90.035003,Thyrhaug2018a,Maly2016a,Chenu2015a}
and others \citep{Sjakste2018,Flesch2008,Lombardo2013,Yu2008,Ruostekoski1998}.
Prevalent theoretical description is given in terms of system- bath
model in constant temperature bath conditions \citep{Breuer2006,Weiss2012},
where the \textit{system }degrees of freedom are coupled to the bath-induced
thermal fluctuations representing the environment, \textit{e.g.},
phonons or vibrational motion of surrounding molecules. Fluctuations
are modeled by an infinite number of quantum harmonic oscillators\textit{\emph{
constituting }}the\textit{ quantum bath} at thermal equilibrium.

These conditions can be fulfilled using the reduced density matrix
approach \citep{Valkunasa,Mukamel1995}. First perturbation order,
with respect to the system-bath coupling, leads to the reduced equations
of motion of the system-only variables, while the bath is averaged
out. Then the system variables indirectly depend on the bath degrees
of freedom via fluctuation correlation functions, which are well-behaved
analytical functions. At the second perturbation order \citep{Breuer2006,Valkunasa},
equations of motion are reminiscent of the Pauli master equation with
relaxation coefficients calculated with respect to the thermal equilibrium.
However, now the resulting equations can lead to unphysical results,
\textit{e.g.,} negative probabilities \citep{Montoya-Castillo2015}.
The more complicated fourth order equations of motion include divergent
parameters and are often avoided \citep{Jang2002}. Non-perturbative,
numerically exact approach of hierarchical equations of motion for
the exponential fluctuation correlation functions is available to
obtain exact dynamics \citep{Tanimura1990a,Tanimura2006,Xu2007},
chain-mapping techniques together with the time-dependent density
matrix renormalization group are alternatively possible for structured
environments \citep{Tamascelli2019,Prior2010}. However, computational
costs limit these methods to models with just few degrees of freedom.
A well known method of stochastic Schrödinger equation requires averaging
over many entangled trajectories to obtain dynamics at finite temperature
\citep{Abramavicius2014,PhysRevB.80.212303,Biele_2012,Diosi1997,DeVega2005,Link2017}.
Its hierarchical realisation \citep{Hartmann2017} improves convergence,
meanwhile, thermofield dynamics approach tries to directly compute
thermally averaged dynamics by mapping the initial thermal density
matrix onto a fictitious bath vacuum state and then coupling system
to it \citep{Reddy2015,Ritschel2015,Borrelli2016a,Chen2017}. Alternativelly,
dissipative dynamics can be obtained by straightforward addition of
a linear friction coefficient to the model Hamiltonian \citep{Martinazzo2006a},
however, it only applies at zero temperature. Yet, in all these cases,
thermal state of the nearest surrounding is not under control.

An important aspect of the bath, more explicitly, of the finite number
of bath oscillators, is its heat capacity. For a single quantum harmonic
oscillator the heat capacity in the limit of weak system-bath coupling
is given by
\begin{align}
c\left(\beta^{-1}\right)= & (\beta\omega)^{2}\frac{\exp(\beta\omega)}{(\exp(\beta\omega)-1)^{2}}\ ,\label{eq:specific heat}
\end{align}
here throughout the paper $\hbar=1$, $\beta=\left(k_{\text{B}}T\right)^{-1}$
is an inverse temperature, $\omega$ is the oscillator frequency.
When the system exchanges energy with a bath made of such oscillators,
its temperature may be affected. If the system-bath energy exchange
is excessively large, the thermal energy can accumulate in the bath
oscillators and this will effectively change thermostat temperature
\citep{Abramavicius2018c}. In most cases, the bath heating effect
is undesirable as, in the system-bath models, the bath is generally
supposed to represent a constant temperature thermostat.

On the other hand, the bath heating effect could be related to the
natural phenomenon of molecular local heating \citep{Chen1993,Ichikawa2007},
\textit{i.e.}, if a molecule quickly dissipates a large amount of
thermal energy to its environment, \textit{e.g., }due to exciton-exciton
annihilation \citep{Gulbinas1996,Valkunas1997,Gulbinas2006} or ultrafast
molecular internal-conversion \citep{Balevicius2018,Balevicius2019},
the local heating of the molecule nearest surrounding takes place
and the further cooling process, the quantum thermalization \citep{Choi2019,Scarani2002},
becomes an important ingredient to consider when describing the corresponding
experiments.

In this work, we introduce the thermalization algorithm to the time
dependent variational theory that allows explicit control over the
bath temperature. By varying the bath size and the thermalization
rate, both the degree of bath heating and the cooling time can be
adjusted. These properties allow to mimic realistic physical conditions,
making presented approach superior to the density operator based approaches,
where the bath heating is excluded, and to the explicit bath models,
where the bath temperature is not controlled.

\section{Fluctuating exciton model}

We consider a molecular aggregate made of $N$ coupled chromophores
at specific \textit{sites}. In the simplest case, the sites represent
distinct molecules that can be electronically excited by, \textit{e.g.},
laser or sunlight irradiation in visible spectral region. Vibrational
normal modes of molecules and of the surrounding medium will be treated
as the baths of harmonic oscillators. Each chromophore is directly
affected only by its own intramolecular vibrations and of its closest
environment, therefore, a separate and uncorrelated (\textit{local})
manifold of vibrational modes $q=1,2,\ldots,Q$ is associated with
each chromophore. Such model is characterized by a Hamiltonian $\hat{H}=\hat{H}_{\text{S}}+\hat{H}_{\text{B}}+\hat{H}_{\text{SB}}$,
with system, bath and system-bath coupling terms being
\begin{align}
\hat{H}_{\text{S}}= & \sum_{n}\varepsilon_{n}\hat{a}_{n}^{\dagger}\hat{a}_{n}+\sum_{n,m}^{n\neq m}J_{nm}\hat{a}_{n}^{\dagger}\hat{a}_{m}\ ,\label{eq:Hamil-S}\\
\hat{H}_{\text{B}}= & \sum_{n,q}\omega_{nq}\hat{b}_{nq}^{\dagger}\hat{b}_{nq}\ ,\label{eq:Hamil-B}\\
\hat{H}_{\text{SB}}= & -\sum_{n}\hat{a}_{n}^{\dagger}\hat{a}_{n}\sum_{q}\omega_{nq}g_{nq}\left(\hat{b}_{nq}^{\dagger}+\hat{b}_{nq}\right)\ ,\label{eq:Hamil-SB}
\end{align}
 Here $\varepsilon_{n}$ denotes $n$th chromophore electronic excitation
energy $J_{nm}$ is the resonant coupling between $n$th and $m$th
chromophores, while $\hat{a}_{n}^{\dagger}$ and $\hat{a}_{n}$ are
the corresponding electronic excitation creation and annihilation
bosonic operators. Frequency of the $q$th vibrational mode in the
$n$th bath is $\omega_{nq}$, the electron-vibrational coupling is
characterized by $g_{nq}$, while $\hat{b}_{nq}^{\dagger}$ and $\hat{b}_{nq}$
are the creation and annihilation bosonic operators of the $q$th
mode in the $n$th bath.

In the following we consider only single electronic excitation in
the aggregate. Time evolution of a non-equilibrium state is described
by Davydov $\text{D}_{2}$ wavefunction \citep{Frenkel1936,Sun2010a}
\begin{equation}
|\Psi_{\text{D}_{2}}\rangle=\sum_{n}^{N}\alpha_{n}\left(t\right)\hat{a}_{n}^{\dagger}\hat{a}_{n}|0\rangle_{\text{el}}\times\prod_{m,q}^{N,Q}|\lambda_{mq}\left(t\right)\rangle\ ,\label{eq:Dav-D2}
\end{equation}
where $\alpha_{n}\left(t\right)$ is the electronic excitation amplitude,
$|0\rangle_{\text{el}}=\prod_{n}|0\rangle_{n}$ is the global ground
state, when all sites are in their electronic ground states. $|\lambda_{mq}\left(t\right)\rangle$
is the coherent state of the $q$th mode in the $m$th bath \citep{Zhang1990,Kais1990}.
It is fully described by the time-dependent complex displacements,
$\lambda_{mq}\left(t\right)$. The time dependent Dirac-Frenkel variational
method allows to obtain equations of motion for parameters $\alpha_{n}$,
$\lambda_{mq}$ \citep{Chorosajev2014a,Jakucionis2018a,Chen2018,Yan2020}
\begin{align}
 & \frac{\text{d}\alpha_{n}\left(t\right)}{\text{dt}}=-\text{i}\alpha_{n}\left(t\right)\varepsilon_{n}-\text{i}\sum_{m}^{m\neq n}\alpha_{m}\left(t\right)J_{nm}\nonumber \\
 & \qquad+\text{i}\alpha_{n}\left(t\right)\sum_{q}\omega_{nq}\left(2g_{nq}-h_{q}\right)\text{Re}\left(\lambda_{nq}\left(t\right)\right)\ ,\label{eq:amp_a}\\
 & \frac{\text{d}\lambda_{mq}\left(t\right)}{\text{dt}}=-\text{i}\omega_{mq}\left(\lambda_{mq}\left(t\right)-h_{q}\left(t\right)\right)\ .\label{eq:amp_l}
\end{align}
Here $h_{q}\left(t\right)=\sum_{i}^{N}g_{iq}\left|\alpha_{i}\left(t\right)\right|^{2}$
is the site population-weighted electron-vibrational coupling strength.
First line in Eq. (\ref{eq:amp_a}) describes dynamics of an isolated
system. Accordingly, the first term on the right hand side of Eq.
(\ref{eq:amp_l}) describes isolated oscillators. Other terms are
due to the system-bath interaction.

Description of the model at the given temperature $T$ requires creation
of statistical ensemble. This is achieved by Monte Carlo sampling
over a statistical thermal ensemble, \textit{i. e.}, over initial
bath oscillator displacements $\lambda_{mq}\left(0\right)$, sampled
from the Glauber-Sudarshan probability distribution \citep{Glauber1963}
\begin{equation}
\mathcal{P}\left(\lambda_{mq}\right)=\mathcal{Z}^{-1}\exp\left(-\left|\lambda_{mq}\right|^{2}\left[\text{e}^{\beta\omega_{mq}}-1\right]\right)\ .\label{eq:Sudarshan-P}
\end{equation}
 The ensemble describes canonical statistics of quantum harmonic oscillators,
which applies to our model prior to external perturbations. The ensemble
averaged quantities will be denoted by $\left\langle \cdots\right\rangle _{\text{th}}$.
The ensemble of exciton trajectories allows to describe irreversible
excitation energy relaxation. While the initial thermal state before
excitation can be properly defined, the bath accepts energy during
exciton relaxation and the state of the bath \emph{after} relaxation
is away from equilibrium. Equations of motion guarantee energy conservation,
hence the combined system-bath cannot thermalize. In order to thermalize
the bath, we extend the original model by introducing the \textit{secondary}
bath (we will refer to the \textit{local }baths as the \textit{primary}
baths). Effective heat capacity of the secondary bath is infinite,
hence, the bath can be characterized by a constant temperature, $T_{\infty}$.
The secondary bath will not be treated explicitly: modes of the primary
baths interact with the secondary bath via \textit{stochastic scattering
events}, or quantum jumps \citep{Plenio1998,Luoma2020}, which affect
the kinetic energy of primary baths modes.

The scattering statistics follows Poisson distribution $P_{mq}\left(\theta,\tau\right)=\frac{1}{\theta!}\left(\tau\nu_{mq}\right)^{\theta}\text{e}^{-\tau\nu_{mq}}$,
which defines the probability of observing $\theta$ scattering events
per time interval $\tau$ with individual event scattering rate $\nu_{mq}$.
Poisson statistics is obtained by simulating a discrete-time Bernoulli
process \citep{Kampen2007,Bertsekas2008} in a limit of $\tau\rightarrow0$
and $\nu_{mq}\tau\ll1$. This is realized in simulations by dividing
the total evolution time $t_{\text{total}}$ into equidistant length
$\tau$ intervals. At the end of each interval, for each mode in the
primary bath, we flip a biased coin with probability $\nu_{mq}\tau$
of landing \textit{``heads}''. If the coin lands \textit{heads,}
we shift momentum of the mode $p_{mq}\left(kt\right)=\sqrt{2}\text{Im}\lambda_{mq}\left(kt\right)$
to a value drawn from the Glauber-Sudarshan distribution, see Eq.
(\ref{eq:Sudarshan-P}), while the coordinate remains unchanged. Otherwise,
if coin lands \textit{tails}, no changes are done. To obtain converged
statistics, we apply the thermalization algorithm to every trajectory
of the thermal ensemble.

\section{Simulation results}

We first demonstrate control of the primary bath temperature of the
simplest possible system, a single, $N=1$, chromophore unit. For
demonstration we set up artificial conditions: the initial primary
bath temperature $T_{1}\left(0\right)=300\ \text{K}$, the secondary
bath is at $T_{\infty}=200\ \text{K}$. The primary bath consists
of $Q=750$ vibrational modes with frequencies, $\omega_{q}=\omega_{0}+\left(q-1\right)\Delta\omega$.
An offset by $\omega_{0}=0.01\ \text{cm}^{-1}$ is introduced for
stability and a step size $\Delta\omega=1\ \text{cm}^{-1}$. The coupling
parameters $g_{nq}$ follow the super-Ohmic spectral density function
$C^{"}\left(\omega\right)=\omega^{s}\exp\left(-\omega/\omega_{c}\right)$
with $s=2$, $\omega_{\text{c}}=100\ \cm$ \citep{Kell2013a,Jakucionis2018a}.
The number of modes and discretization parameters are sufficient to
obtain convergent model dynamics. For thermalization, we consider
scattering rates of all modes to be equal $\nu_{mq}\rightarrow\nu$,
and the scattering step size $\tau=0.01\ \text{ps}^{-1}$. Thermal
ensemble consists of $5000$ trajectories.

\begin{figure}
\includegraphics[width=8.6cm]{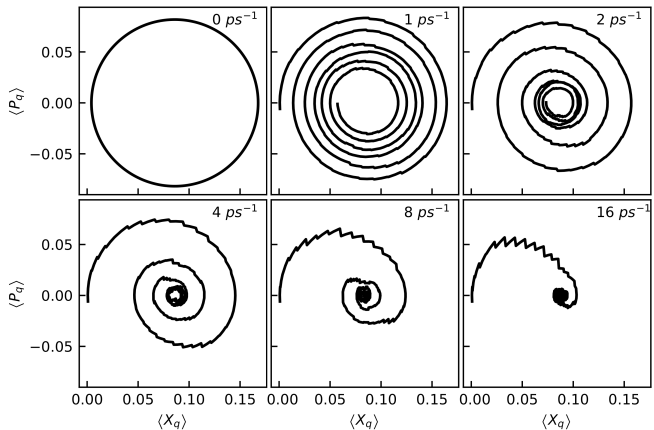}

\caption{Phase space trajectory of one specific bath mode $\omega_{q}=100\ \protect\cm$
for a single excited chromophore calculated with various scattering
rates $\nu$. Initial temperature of the primary bath is $T_{1}\left(0\right)=300\ \text{K}$
and the secondary bath is at a constant temperature $T_{\infty}=200\ \text{K}$.
Scattering step size is $\tau=0.01\ \text{ps}$. \label{fig:Phase-trajectories}
Wiggles in dynamics are due to finite size ensemble averaging (5000
trajectories).}
\end{figure}

In Fig. (\ref{fig:Phase-trajectories}) the coordinate $\left\langle x_{1q}\right\rangle _{\text{th}}=\left\langle \sqrt{2}\text{Re}\lambda_{1q}\right\rangle _{\text{th}}$
and momentum $\left\langle p_{1q}\right\rangle _{\text{th}}=\left\langle \sqrt{2}\text{Im}\lambda_{mq}\right\rangle _{\text{th}}$
phase space trajectory of a single $100\ \text{cm}^{-1}$ frequency
vibrational mode, calculated with various scattering rates $\nu$
is presented. The oscillator, in the absence of thermalization, evolves
along a closed trajectory around $x_{1q}^{\text{min}}=\sqrt{2}g_{1q}$.
Applying thermalization procedure, a dissipative type trajectory is
observed. The coordinate $\left\langle x_{1q}\right\rangle _{\text{th}}$
equilibrates to $x_{1q}^{\text{min}}$ (equilibrium is shifted from
zero due to coupling with the system), while momentum $\left\langle p_{1q}\right\rangle _{\text{th}}$
approaches zero. The thermalization time can be adjusted by changing
the scattering rate, $\nu$. Both weakly damped and overdamped regimes
become available.

\begin{figure}
\includegraphics[width=8.6cm]{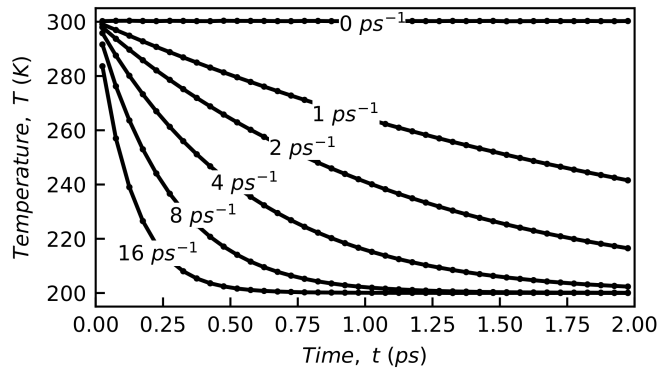}

\caption{The primary bath temperature $T_{1}\left(t\right)$ calculated with
various scattering rates $\nu$. Initial temperature of the primary
bath is $T_{1}\left(0\right)=300\ \text{K}$ and the secondary bath
is at a constant temperature $T_{\infty}=200\ \text{K}$. Scattering
step size is $\tau=0.01\ \text{ps}$.\label{fig:bath-temp}}
\end{figure}

Transient temperature of the primary bath can be estimated \citep{Abramavicius2018c}
by computing the average kinetic energy $\left\langle K_{mq}\left(t,\epsilon\right)\right\rangle _{\text{th}}$
over time interval $\epsilon$. The parameter $\epsilon$ then implies
the resolution. For the whole primary bath the transient temperature
is then given by 
\begin{equation}
T_{m}\left(t\right)=\frac{1}{k_{\text{B}}Q}\sum_{q=1}^{Q}\omega_{mq}\ln\left(1+\frac{\omega_{mq}}{2\left\langle K_{mq}\left(t,\epsilon\right)\right\rangle _{\text{th}}}\right)^{-1}\ .
\end{equation}
 In Fig. (\ref{fig:bath-temp}) we present the primary bath temperature
calculated with $\epsilon=50\ \text{fs}$ and various scattering rates,
$\nu$. In the absence of thermalization, the primary bath temperature
remains at the initial value of $T_{1}\left(0\right)=300\ \text{K}$.
Meanwhile, thermalization introduces cooling of the primary bath down
to the temperature of the secondary bath. The scattering rate, $\nu$,
allows to control the thermalization time.

\begin{figure}
\includegraphics[width=8.6cm]{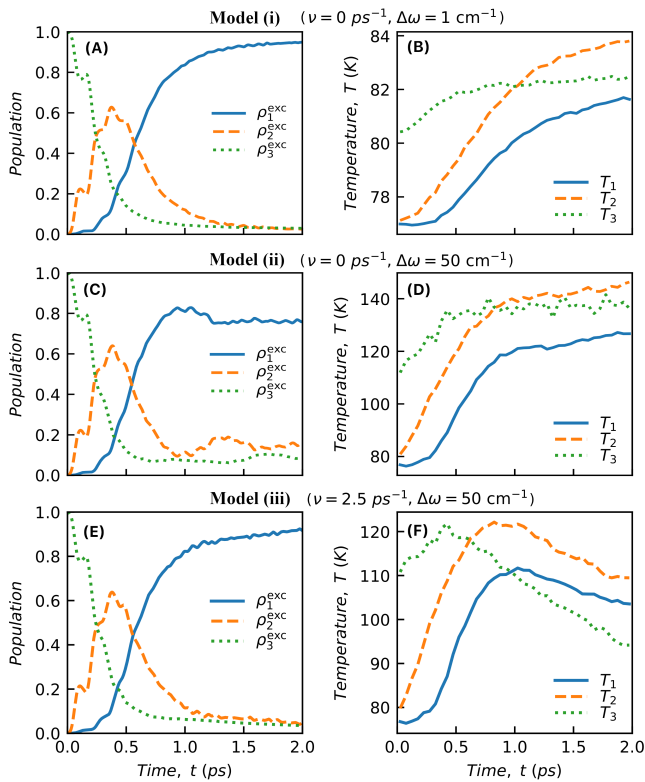}\caption{Multi site-bath model exciton state populations $\rho_{n}^{\text{exc}}\left(t\right)$
and local bath temperatures $T_{m}\left(t\right)$ calculated in (i)
the \textit{dense} primary bath discretization regime without thermalization,
(ii) the \textit{sparse} discretization regime without thermalization
and (iii) the sparcely discretized bath with thermalization ($\nu=2.5\ \text{ps}^{-1}$).
\label{fig:Multi-site-P-T}}
\end{figure}

The temperature control and stability considerably affects the electronic
excitation dynamics. To demonstrate the sensitivity of the excitation
evolution to the thermalization we consider a linear $N=3$ chromophore
aggregate, with chromophore transition energies $0,\ 250,\ 500\ \text{cm}^{-1}$,
and nearest neighbor coupling $J=100\ \text{cm}^{-1}$. \textit{\emph{Excited
states of such chromophore aggregate are}}\textit{ excitons} \citep{Valkunasa,Amerongen2010}.
They represent electronic excitations delocalized over several sites
with time dependent delocalization length \citep{Chorosajev2016c}.
Hence, we switch to the eigenstate basis (exciton representation,
defined by $\hat{H}_{\text{S}}\psi^{(\mathrm{exc})}=\varepsilon\psi^{(\mathrm{exc})}$):
$\rho_{e}^{\text{(exc)}}\left(t\right)=\sum_{n,m}\left(\psi_{ne}^{(\mathrm{exc})}\right)^{\star}\left\langle \alpha_{n}^{\ast}\left(t\right)\alpha_{m}\left(t\right)\right\rangle _{th}\psi_{me}^{(\mathrm{exc})}$.
The initial electronic state correspond to the optically excited highest
energy exciton eigen state. The parameters of the primary baths of
chromophores are the same as above, however, now the initial primary
bath temperature and the secondary bath temperature is the same $T_{m}\left(0\right)=T_{\infty}=77\ \text{K}$.
The thermal ensemble consists of $240$ trajectories. In Fig. (\ref{fig:Multi-site-P-T})
we present exciton state populations $\rho_{e}^{\text{exc}}\left(t\right)$
and the primary bath temperatures $T_{m}\left(t\right)$ calculated
in (i) the \textit{dense} primary bath discretization regime without
thermalization (the bath discretization step size is $\Delta\omega=1\ \text{cm}^{-1}$,
$Q=750$ vibrational modes per site), in (ii) the \textit{sparse}
discretization regime without thermalization ($\Delta\omega=50\ \text{cm}^{-1}$,
$Q=15$) and (iii) the sparce discretized bath with thermalization
($\nu=2.5\ \text{ps}^{-1}$).

Consider the excitation dynamics without thermalization. In models
(i) and (ii) exciton populations sequentially relax to lower energy
exciton states, eventually reaching the lowest energy state. \citep{Moix2012,Subas2012,Gelzinis2020}.
\begin{comment}
For an infinite bath in a weak system-bath coupling limit, $\Lambda_{m}\ll J_{nm}$,
population distribution of the exciton steady state would be a Boltzmann
distribution at a temperature $T_{\infty}$. In our case, system-bath
and site-site coupling are on the same order, therefore steady state
population distribution

While distribution of a dense regime shows qualitative agreement with
Boltzmann distribution, and, thus, can be considered as an approximation
to an ideal thermostat, distribution of a sparse regime is truly non-Boltzmann.
\end{comment}
The final population distribution in the sparse regime, model (ii),
significantly differs from the dense case. Origin of the discrepancy
is two-fold: the bath recursion time $t_{\text{rec}}=2\pi/\Delta\omega$
for model (ii) is shorter than the calculation time $t_{\text{rec}}<t_{\text{total}}$,
and the sparse primary bath shows significant growth of the bath temperature,
compare Fig. (\ref{fig:Multi-site-P-T}B and \ref{fig:Multi-site-P-T}D).
Both of these drawbacks are addressed by introducing the bath thermalization
in model (iii). Looking at Fig. (\ref{fig:Multi-site-P-T}E), we see
that the exciton population dynamics and steady state values for model
(iii) become quantitatively comparable to case of model (i).

\section{Discussion}

A single quantum harmonic oscillator is characterized by a specific
heat $c(\beta^{-1})<k_{\text{B}}$, which depends on temperature as
given by Eq. (\ref{eq:specific heat}). For a given set of bath oscillators
the specific heat at a given temperature can be estimated, however,
the harmonic oscillators of the bath as defined by Eq. (\ref{eq:Hamil-B})
do not exchange energy. Accordingly, as the system relaxes, only a
few in-resonant oscillators accept the energy and diverges away from
equilibrium \citep{Chin2013a}. Hence, the temperature at which excitation
dynamics occur no longer match the initial bath temperature - local
heating takes place.

A straightforward approach to avoid heating is to increase the bath
density of states  until dynamics of interest converges (in our model,
this is achieved by increasing the number of bath oscillators). However,
this is acceptable only for small systems, since computation effort
scales quadratically with both number of sites and bath oscillators.
Thermalization can be utilized to steer the bath to the required temperature.
Additional merit of thermalization is the significant reduction of
the the number of vibrational modes needed per bath. Our simulations
show convergence with just $15$ modes per bath while maintaining
comparable exciton relaxation dynamics (Fig. \ref{fig:Multi-site-P-T}).

In effort to reduce the computational effort, Wang \textit{et al.}
have used \citep{Wang2016a} a logarithmic bath discretization. However,
high frequency representation of the continuous spectral density becomes
poor. Our model is in line with explicit surrogate Hamiltonian \citep{Baer1997}
and its stochastic realization \citep{Katz2008,Torrontegui2016a,Habecker2019a},
while our approach does not require the explicit modeling of the secondary
bath, it still maintaining proper quantum dynamics in the system.

The time-dependent variational approach with Davydov $\text{D}_{2}$
ansatz can be improved by considering more complex Davydov $\text{ans\ensuremath{\ddot{a}}tze}$
family members, \textit{e.g.}, multitude of $\text{D}_{1}$ ansatz
(multi-$\text{D}_{1}$) and multi-$\text{D}_{2}$ \citep{Zhou2015b,Wang2016a,Zhou2016a,Chen2018}
or its Born-Oppenheimer approximated variant \citep{Jakucionis2020},
$\text{sD}_{2}$. In either way, they all suffer from finite bath
heating capacity, in most cases, even stronger than the $\text{D}_{2}$
ansatz, because of significantly increased computational effort needed
to propagate numerous bath oscillators. Work is in progress on adapting
the presented thermalization algorithm to these more intricate $\text{ans\ensuremath{\ddot{a}}tze}$.

In conclusion, we present a system-bath model with stochastic bath
thermalization using the time-dependent variational approach with
Davydov $\text{D}_{2}$ ansatz. Thermalization allows to steer bath
vibrational modes evolution towards equilibrium thermal state of selected
temperature in a controlled way, and at the same time, for the bath
to still maintain an aspect of being coupled to the system. In addition,
by analysing exciton relaxation dynamics of a chromophore aggregate
with thermalization, we found the exciton dynamics to converge with
much smaller number of bath modes, significantly speeding up numerical
computation.
\begin{acknowledgments}
We thank the Research Council of Lithuania for financial support (grant
No: SMIP-20-47). Computations were performed on resources at the High
Performance Computing Center, \textquoteleft \textquoteleft HPC Sauletekis\textquoteright \textquoteright{}
in Vilnius University Faculty of Physics.
\end{acknowledgments}

\bibliographystyle{apsrev4-2}
\addcontentsline{toc}{section}{\refname}\bibliography{7ring_prl_v2,7ring,7ring_pra}

\begin{thebibliography}{74}%
\makeatletter
\providecommand \@ifxundefined [1]{%
 \@ifx{#1\undefined}
}%
\providecommand \@ifnum [1]{%
 \ifnum #1\expandafter \@firstoftwo
 \else \expandafter \@secondoftwo
 \fi
}%
\providecommand \@ifx [1]{%
 \ifx #1\expandafter \@firstoftwo
 \else \expandafter \@secondoftwo
 \fi
}%
\providecommand \natexlab [1]{#1}%
\providecommand \enquote  [1]{``#1''}%
\providecommand \bibnamefont  [1]{#1}%
\providecommand \bibfnamefont [1]{#1}%
\providecommand \citenamefont [1]{#1}%
\providecommand \href@noop [0]{\@secondoftwo}%
\providecommand \href [0]{\begingroup \@sanitize@url \@href}%
\providecommand \@href[1]{\@@startlink{#1}\@@href}%
\providecommand \@@href[1]{\endgroup#1\@@endlink}%
\providecommand \@sanitize@url [0]{\catcode `\\12\catcode `\$12\catcode
  `\&12\catcode `\#12\catcode `\^12\catcode `\_12\catcode `\%12\relax}%
\providecommand \@@startlink[1]{}%
\providecommand \@@endlink[0]{}%
\providecommand \url  [0]{\begingroup\@sanitize@url \@url }%
\providecommand \@url [1]{\endgroup\@href {#1}{\urlprefix }}%
\providecommand \urlprefix  [0]{URL }%
\providecommand \Eprint [0]{\href }%
\providecommand \doibase [0]{https://doi.org/}%
\providecommand \selectlanguage [0]{\@gobble}%
\providecommand \bibinfo  [0]{\@secondoftwo}%
\providecommand \bibfield  [0]{\@secondoftwo}%
\providecommand \translation [1]{[#1]}%
\providecommand \BibitemOpen [0]{}%
\providecommand \bibitemStop [0]{}%
\providecommand \bibitemNoStop [0]{.\EOS\space}%
\providecommand \EOS [0]{\spacefactor3000\relax}%
\providecommand \BibitemShut  [1]{\csname bibitem#1\endcsname}%
\let\auto@bib@innerbib\@empty
%</preamble>
\bibitem [{\citenamefont {Dorfman}\ \emph {et~al.}(2016)\citenamefont
  {Dorfman}, \citenamefont {Schlawin},\ and\ \citenamefont
  {Mukamel}}]{Dorfman2016}%
  \BibitemOpen
  \bibfield  {author} {\bibinfo {author} {\bibfnamefont {K.~E.}\ \bibnamefont
  {Dorfman}}, \bibinfo {author} {\bibfnamefont {F.}~\bibnamefont {Schlawin}},\
  and\ \bibinfo {author} {\bibfnamefont {S.}~\bibnamefont {Mukamel}},\ }\href
  {https://doi.org/10.1103/RevModPhys.88.045008} {\bibfield  {journal}
  {\bibinfo  {journal} {Rev. Mod. Phys.}\ }\textbf {\bibinfo {volume} {88}},\
  \bibinfo {pages} {45008} (\bibinfo {year} {2016})}\BibitemShut {NoStop}%
\bibitem [{\citenamefont {Mukamel}(1995)}]{Mukamel1995}%
  \BibitemOpen
  \bibfield  {author} {\bibinfo {author} {\bibfnamefont {S.~S.}\ \bibnamefont
  {Mukamel}},\ }\href@noop {} {\emph {\bibinfo {title} {{Principles of
  nonlinear optical spectroscopy}}}}\ (\bibinfo  {publisher} {Oxford University
  Press},\ \bibinfo {year} {1995})\ p.\ \bibinfo {pages} {543}\BibitemShut
  {NoStop}%
\bibitem [{\citenamefont {Jaku{\v{c}}ionis}\ \emph {et~al.}(2018)\citenamefont
  {Jaku{\v{c}}ionis}, \citenamefont {Choro{\v{s}}ajev},\ and\ \citenamefont
  {Abramavi{\v{c}}ius}}]{Jakucionis2018a}%
  \BibitemOpen
  \bibfield  {author} {\bibinfo {author} {\bibfnamefont {M.}~\bibnamefont
  {Jaku{\v{c}}ionis}}, \bibinfo {author} {\bibfnamefont {V.}~\bibnamefont
  {Choro{\v{s}}ajev}},\ and\ \bibinfo {author} {\bibfnamefont {D.}~\bibnamefont
  {Abramavi{\v{c}}ius}},\ }\href
  {https://doi.org/10.1016/j.chemphys.2018.07.018} {\bibfield  {journal}
  {\bibinfo  {journal} {Chem. Phys.}\ }\textbf {\bibinfo {volume} {515}},\
  \bibinfo {pages} {193} (\bibinfo {year} {2018})}\BibitemShut {NoStop}%
\bibitem [{\citenamefont {Choro{\v{s}}ajev}\ \emph {et~al.}(2017)\citenamefont
  {Choro{\v{s}}ajev}, \citenamefont {Mar{\v{c}}iulionis},\ and\ \citenamefont
  {Abramavicius}}]{Chorosajev2017a}%
  \BibitemOpen
  \bibfield  {author} {\bibinfo {author} {\bibfnamefont {V.}~\bibnamefont
  {Choro{\v{s}}ajev}}, \bibinfo {author} {\bibfnamefont {T.}~\bibnamefont
  {Mar{\v{c}}iulionis}},\ and\ \bibinfo {author} {\bibfnamefont
  {D.}~\bibnamefont {Abramavicius}},\ }\href
  {https://doi.org/10.1063/1.4985910} {\bibfield  {journal} {\bibinfo
  {journal} {J. Chem. Phys.}\ }\textbf {\bibinfo {volume} {147}},\ \bibinfo
  {pages} {74114} (\bibinfo {year} {2017})}\BibitemShut {NoStop}%
\bibitem [{\citenamefont {Schr{\"{o}}ter}\ \emph {et~al.}(2015)\citenamefont
  {Schr{\"{o}}ter}, \citenamefont {Ivanov}, \citenamefont {Schulze},
  \citenamefont {Polyutov}, \citenamefont {Yan}, \citenamefont {Pullerits},\
  and\ \citenamefont {K{\"{u}}hn}}]{Schroter2015}%
  \BibitemOpen
  \bibfield  {author} {\bibinfo {author} {\bibfnamefont {M.}~\bibnamefont
  {Schr{\"{o}}ter}}, \bibinfo {author} {\bibfnamefont {S.}~\bibnamefont
  {Ivanov}}, \bibinfo {author} {\bibfnamefont {J.}~\bibnamefont {Schulze}},
  \bibinfo {author} {\bibfnamefont {S.}~\bibnamefont {Polyutov}}, \bibinfo
  {author} {\bibfnamefont {Y.}~\bibnamefont {Yan}}, \bibinfo {author}
  {\bibfnamefont {T.}~\bibnamefont {Pullerits}},\ and\ \bibinfo {author}
  {\bibfnamefont {O.}~\bibnamefont {K{\"{u}}hn}},\ }\href
  {https://doi.org/10.1016/j.physrep.2014.12.001} {\bibfield  {journal}
  {\bibinfo  {journal} {Phys. Rep.}\ }\textbf {\bibinfo {volume} {567}},\
  \bibinfo {pages} {1} (\bibinfo {year} {2015})}\BibitemShut {NoStop}%
\bibitem [{\citenamefont {Choro{\v{s}}ajev}\ \emph {et~al.}(2014)\citenamefont
  {Choro{\v{s}}ajev}, \citenamefont {Gelzinis}, \citenamefont {Valkunas},\ and\
  \citenamefont {Abramavicius}}]{Chorosajev2014a}%
  \BibitemOpen
  \bibfield  {author} {\bibinfo {author} {\bibfnamefont {V.}~\bibnamefont
  {Choro{\v{s}}ajev}}, \bibinfo {author} {\bibfnamefont {A.}~\bibnamefont
  {Gelzinis}}, \bibinfo {author} {\bibfnamefont {L.}~\bibnamefont {Valkunas}},\
  and\ \bibinfo {author} {\bibfnamefont {D.}~\bibnamefont {Abramavicius}},\
  }\href {https://doi.org/10.1063/1.4884275} {\bibfield  {journal} {\bibinfo
  {journal} {J. Chem. Phys.}\ }\textbf {\bibinfo {volume} {140}},\ \bibinfo
  {pages} {244108} (\bibinfo {year} {2014})}\BibitemShut {NoStop}%
\bibitem [{\citenamefont {Valkunas}\ \emph {et~al.}(2013)\citenamefont
  {Valkunas}, \citenamefont {Abramavicius},\ and\ \citenamefont
  {Man{\v{c}}al}}]{Valkunasa}%
  \BibitemOpen
  \bibfield  {author} {\bibinfo {author} {\bibfnamefont {L.}~\bibnamefont
  {Valkunas}}, \bibinfo {author} {\bibfnamefont {D.}~\bibnamefont
  {Abramavicius}},\ and\ \bibinfo {author} {\bibfnamefont {T.}~\bibnamefont
  {Man{\v{c}}al}},\ }\href
  {https://books.google.lt/books?hl=lt{\&}lr={\&}id=ZR5AAQAAQBAJ{\&}oi=fnd{\&}pg=PT11{\&}ots=4dIsblNNm{\_}{\&}sig=0dXw7Xp6xvhGcZoBxiDa8huh22k{\&}redir{\_}esc=y{\#}v=onepage{\&}q{\&}f=false}
  {\emph {\bibinfo {title} {{Molecular Excitation Dynamics and Relaxation}}}}\
  (\bibinfo  {publisher} {Wiley-VCH},\ \bibinfo {year} {2013})\BibitemShut
  {NoStop}%
\bibitem [{\citenamefont {May}\ and\ \citenamefont
  {K{\"{u}}hn}(2011)}]{May2011}%
  \BibitemOpen
  \bibfield  {author} {\bibinfo {author} {\bibfnamefont {V.}~\bibnamefont
  {May}}\ and\ \bibinfo {author} {\bibfnamefont {O.}~\bibnamefont
  {K{\"{u}}hn}},\ }\href {https://doi.org/10.1002/9783527633791} {\emph
  {\bibinfo {title} {{Charge and Energy Transfer Dynamics in Molecular
  Systems}}}}\ (\bibinfo  {publisher} {Wiley-VCH Verlag GmbH {\&} Co. KGaA},\
  \bibinfo {address} {Weinheim, Germany},\ \bibinfo {year} {2011})\ p.\
  \bibinfo {pages} {562}\BibitemShut {NoStop}%
\bibitem [{\citenamefont {Jang}\ and\ \citenamefont
  {Mennucci}(2018)}]{RevModPhys.90.035003}%
  \BibitemOpen
  \bibfield  {author} {\bibinfo {author} {\bibfnamefont {S.~J.}\ \bibnamefont
  {Jang}}\ and\ \bibinfo {author} {\bibfnamefont {B.}~\bibnamefont
  {Mennucci}},\ }\href {https://doi.org/10.1103/RevModPhys.90.035003}
  {\bibfield  {journal} {\bibinfo  {journal} {Rev. Mod. Phys.}\ }\textbf
  {\bibinfo {volume} {90}},\ \bibinfo {pages} {35003} (\bibinfo {year}
  {2018})}\BibitemShut {NoStop}%
\bibitem [{\citenamefont {Thyrhaug}\ \emph {et~al.}(2018)\citenamefont
  {Thyrhaug}, \citenamefont {Lincoln}, \citenamefont {Branchi}, \citenamefont
  {Cerullo}, \citenamefont {Perl{\'{i}}k}, \citenamefont {{\v{S}}anda},
  \citenamefont {Lokstein},\ and\ \citenamefont {Hauer}}]{Thyrhaug2018a}%
  \BibitemOpen
  \bibfield  {author} {\bibinfo {author} {\bibfnamefont {E.}~\bibnamefont
  {Thyrhaug}}, \bibinfo {author} {\bibfnamefont {C.~N.}\ \bibnamefont
  {Lincoln}}, \bibinfo {author} {\bibfnamefont {F.}~\bibnamefont {Branchi}},
  \bibinfo {author} {\bibfnamefont {G.}~\bibnamefont {Cerullo}}, \bibinfo
  {author} {\bibfnamefont {V.}~\bibnamefont {Perl{\'{i}}k}}, \bibinfo {author}
  {\bibfnamefont {F.}~\bibnamefont {{\v{S}}anda}}, \bibinfo {author}
  {\bibfnamefont {H.}~\bibnamefont {Lokstein}},\ and\ \bibinfo {author}
  {\bibfnamefont {J.}~\bibnamefont {Hauer}},\ }\href
  {https://doi.org/10.1007/s11120-017-0398-3} {\bibfield  {journal} {\bibinfo
  {journal} {Photosynth. Res.}\ }\textbf {\bibinfo {volume} {135}},\ \bibinfo
  {pages} {45} (\bibinfo {year} {2018})}\BibitemShut {NoStop}%
\bibitem [{\citenamefont {Mal{\'{y}}}\ \emph {et~al.}(2016)\citenamefont
  {Mal{\'{y}}}, \citenamefont {Gruber}, \citenamefont {Cogdell}, \citenamefont
  {Man{\v{c}}al},\ and\ \citenamefont {{Van Grondelle}}}]{Maly2016a}%
  \BibitemOpen
  \bibfield  {author} {\bibinfo {author} {\bibfnamefont {P.}~\bibnamefont
  {Mal{\'{y}}}}, \bibinfo {author} {\bibfnamefont {J.~M.}\ \bibnamefont
  {Gruber}}, \bibinfo {author} {\bibfnamefont {R.~J.}\ \bibnamefont {Cogdell}},
  \bibinfo {author} {\bibfnamefont {T.}~\bibnamefont {Man{\v{c}}al}},\ and\
  \bibinfo {author} {\bibfnamefont {R.}~\bibnamefont {{Van Grondelle}}},\
  }\href {https://doi.org/10.1073/pnas.1522265113} {\bibfield  {journal}
  {\bibinfo  {journal} {Proc. Natl. Acad. Sci. U.S.A.}\ }\textbf {\bibinfo
  {volume} {113}},\ \bibinfo {pages} {2934} (\bibinfo {year}
  {2016})}\BibitemShut {NoStop}%
\bibitem [{\citenamefont {Chenu}\ and\ \citenamefont
  {Scholes}(2015)}]{Chenu2015a}%
  \BibitemOpen
  \bibfield  {author} {\bibinfo {author} {\bibfnamefont {A.}~\bibnamefont
  {Chenu}}\ and\ \bibinfo {author} {\bibfnamefont {G.~D.}\ \bibnamefont
  {Scholes}},\ }\href {https://doi.org/10.1146/annurev-physchem-040214-121713}
  {\bibfield  {journal} {\bibinfo  {journal} {Annu. Rev. Phys. Chem.}\ }\textbf
  {\bibinfo {volume} {66}},\ \bibinfo {pages} {69} (\bibinfo {year}
  {2015})}\BibitemShut {NoStop}%
\bibitem [{\citenamefont {Sjakste}\ \emph {et~al.}(2018)\citenamefont
  {Sjakste}, \citenamefont {Tanimura}, \citenamefont {Barbarino}, \citenamefont
  {Perfetti},\ and\ \citenamefont {Vast}}]{Sjakste2018}%
  \BibitemOpen
  \bibfield  {author} {\bibinfo {author} {\bibfnamefont {J.}~\bibnamefont
  {Sjakste}}, \bibinfo {author} {\bibfnamefont {K.}~\bibnamefont {Tanimura}},
  \bibinfo {author} {\bibfnamefont {G.}~\bibnamefont {Barbarino}}, \bibinfo
  {author} {\bibfnamefont {L.}~\bibnamefont {Perfetti}},\ and\ \bibinfo
  {author} {\bibfnamefont {N.}~\bibnamefont {Vast}},\ }\href
  {https://doi.org/10.1088/1361-648X/aad487} {\bibfield  {journal} {\bibinfo
  {journal} {J. Phys. Condens. Matter}\ }\textbf {\bibinfo {volume} {30}},\
  \bibinfo {pages} {353001} (\bibinfo {year} {2018})}\BibitemShut {NoStop}%
\bibitem [{\citenamefont {Flesch}\ \emph {et~al.}(2008)\citenamefont {Flesch},
  \citenamefont {Cramer}, \citenamefont {McCulloch}, \citenamefont
  {Schollw{\"{o}}ck},\ and\ \citenamefont {Eisert}}]{Flesch2008}%
  \BibitemOpen
  \bibfield  {author} {\bibinfo {author} {\bibfnamefont {A.}~\bibnamefont
  {Flesch}}, \bibinfo {author} {\bibfnamefont {M.}~\bibnamefont {Cramer}},
  \bibinfo {author} {\bibfnamefont {I.~P.}\ \bibnamefont {McCulloch}}, \bibinfo
  {author} {\bibfnamefont {U.}~\bibnamefont {Schollw{\"{o}}ck}},\ and\ \bibinfo
  {author} {\bibfnamefont {J.}~\bibnamefont {Eisert}},\ }\href
  {https://doi.org/10.1103/PhysRevA.78.033608} {\bibfield  {journal} {\bibinfo
  {journal} {Phys. Rev. A}\ }\textbf {\bibinfo {volume} {78}},\ \bibinfo
  {pages} {033608} (\bibinfo {year} {2008})}\BibitemShut {NoStop}%
\bibitem [{\citenamefont {Lombardo}\ and\ \citenamefont
  {Turiaci}(2013)}]{Lombardo2013}%
  \BibitemOpen
  \bibfield  {author} {\bibinfo {author} {\bibfnamefont {F.~C.}\ \bibnamefont
  {Lombardo}}\ and\ \bibinfo {author} {\bibfnamefont {G.~J.}\ \bibnamefont
  {Turiaci}},\ }\href {https://doi.org/10.1103/PhysRevD.87.084028} {\bibfield
  {journal} {\bibinfo  {journal} {Phys. Rev. D}\ }\textbf {\bibinfo {volume}
  {87}},\ \bibinfo {pages} {84028} (\bibinfo {year} {2013})}\BibitemShut
  {NoStop}%
\bibitem [{\citenamefont {Yu}\ and\ \citenamefont {Zhang}(2008)}]{Yu2008}%
  \BibitemOpen
  \bibfield  {author} {\bibinfo {author} {\bibfnamefont {H.}~\bibnamefont
  {Yu}}\ and\ \bibinfo {author} {\bibfnamefont {J.}~\bibnamefont {Zhang}},\
  }\href {https://doi.org/10.1103/PhysRevD.77.024031} {\bibfield  {journal}
  {\bibinfo  {journal} {Phys. Rev. D}\ }\textbf {\bibinfo {volume} {77}},\
  \bibinfo {pages} {24031} (\bibinfo {year} {2008})}\BibitemShut {NoStop}%
\bibitem [{\citenamefont {Ruostekoski}\ and\ \citenamefont
  {Walls}(1998)}]{Ruostekoski1998}%
  \BibitemOpen
  \bibfield  {author} {\bibinfo {author} {\bibfnamefont {J.}~\bibnamefont
  {Ruostekoski}}\ and\ \bibinfo {author} {\bibfnamefont {D.~F.}\ \bibnamefont
  {Walls}},\ }\href {https://doi.org/10.1103/PhysRevA.58.R50} {\bibfield
  {journal} {\bibinfo  {journal} {Phys. Rev. A}\ }\textbf {\bibinfo {volume}
  {58}},\ \bibinfo {pages} {R50} (\bibinfo {year} {1998})}\BibitemShut
  {NoStop}%
\bibitem [{\citenamefont {Breuer}\ and\ \citenamefont
  {Petruccione}(2007)}]{Breuer2006}%
  \BibitemOpen
  \bibfield  {author} {\bibinfo {author} {\bibfnamefont {H.-P.}\ \bibnamefont
  {Breuer}}\ and\ \bibinfo {author} {\bibfnamefont {F.}~\bibnamefont
  {Petruccione}},\ }\href
  {https://doi.org/10.1093/acprof:oso/9780199213900.001.0001} {\emph {\bibinfo
  {title} {{The Theory of Open Quantum Systems}}}}\ (\bibinfo  {publisher}
  {Oxford University Press},\ \bibinfo {year} {2007})\ p.\ \bibinfo {pages}
  {636}\BibitemShut {NoStop}%
\bibitem [{\citenamefont {Weiss}(2012)}]{Weiss2012}%
  \BibitemOpen
  \bibfield  {author} {\bibinfo {author} {\bibfnamefont {U.}~\bibnamefont
  {Weiss}},\ }\href {https://doi.org/10.1142/8334} {\emph {\bibinfo {title}
  {{Quantum Dissipative Systems}}}}\ (\bibinfo  {publisher} {World
  Scientific},\ \bibinfo {year} {2012})\BibitemShut {NoStop}%
\bibitem [{\citenamefont {Montoya-Castillo}\ \emph {et~al.}(2015)\citenamefont
  {Montoya-Castillo}, \citenamefont {Berkelbach},\ and\ \citenamefont
  {Reichman}}]{Montoya-Castillo2015}%
  \BibitemOpen
  \bibfield  {author} {\bibinfo {author} {\bibfnamefont {A.}~\bibnamefont
  {Montoya-Castillo}}, \bibinfo {author} {\bibfnamefont {T.~C.}\ \bibnamefont
  {Berkelbach}},\ and\ \bibinfo {author} {\bibfnamefont {D.~R.}\ \bibnamefont
  {Reichman}},\ }\href {https://doi.org/10.1063/1.4935443} {\bibfield
  {journal} {\bibinfo  {journal} {J. Chem. Phys.}\ }\textbf {\bibinfo {volume}
  {143}},\ \bibinfo {pages} {194108} (\bibinfo {year} {2015})}\BibitemShut
  {NoStop}%
\bibitem [{\citenamefont {Jang}\ \emph {et~al.}(2002)\citenamefont {Jang},
  \citenamefont {Cao},\ and\ \citenamefont {Silbey}}]{Jang2002}%
  \BibitemOpen
  \bibfield  {author} {\bibinfo {author} {\bibfnamefont {S.}~\bibnamefont
  {Jang}}, \bibinfo {author} {\bibfnamefont {J.}~\bibnamefont {Cao}},\ and\
  \bibinfo {author} {\bibfnamefont {R.~J.}\ \bibnamefont {Silbey}},\ }\href
  {https://doi.org/10.1063/1.1445105} {\bibfield  {journal} {\bibinfo
  {journal} {J. Chem. Phys.}\ }\textbf {\bibinfo {volume} {116}},\ \bibinfo
  {pages} {2705} (\bibinfo {year} {2002})}\BibitemShut {NoStop}%
\bibitem [{\citenamefont {Tanimura}(1990)}]{Tanimura1990a}%
  \BibitemOpen
  \bibfield  {author} {\bibinfo {author} {\bibfnamefont {Y.}~\bibnamefont
  {Tanimura}},\ }\href {https://doi.org/10.1103/PhysRevA.41.6676} {\bibfield
  {journal} {\bibinfo  {journal} {Phys. Rev. A}\ }\textbf {\bibinfo {volume}
  {41}},\ \bibinfo {pages} {6676} (\bibinfo {year} {1990})}\BibitemShut
  {NoStop}%
\bibitem [{\citenamefont {Tanimura}(2006)}]{Tanimura2006}%
  \BibitemOpen
  \bibfield  {author} {\bibinfo {author} {\bibfnamefont {Y.}~\bibnamefont
  {Tanimura}},\ }\href {https://doi.org/10.1143/JPSJ.75.082001} {\bibfield
  {journal} {\bibinfo  {journal} {J. Phys. Soc. Japan}\ }\textbf {\bibinfo
  {volume} {75}},\ \bibinfo {pages} {82001} (\bibinfo {year}
  {2006})}\BibitemShut {NoStop}%
\bibitem [{\citenamefont {Xu}\ and\ \citenamefont {Yan}(2007)}]{Xu2007}%
  \BibitemOpen
  \bibfield  {author} {\bibinfo {author} {\bibfnamefont {R.-X.}\ \bibnamefont
  {Xu}}\ and\ \bibinfo {author} {\bibfnamefont {Y.J.}~\bibnamefont {Yan}},\
  }\href {https://doi.org/10.1103/PhysRevE.75.031107} {\bibfield  {journal}
  {\bibinfo  {journal} {Phys. Rev. E}\ }\textbf {\bibinfo {volume} {75}},\
  \bibinfo {pages} {031107} (\bibinfo {year} {2007})}\BibitemShut {NoStop}%
\bibitem [{\citenamefont {Tamascelli}\ \emph {et~al.}(2019)\citenamefont
  {Tamascelli}, \citenamefont {Smirne}, \citenamefont {Lim}, \citenamefont
  {Huelga},\ and\ \citenamefont {Plenio}}]{Tamascelli2019}%
  \BibitemOpen
  \bibfield  {author} {\bibinfo {author} {\bibfnamefont {D.}~\bibnamefont
  {Tamascelli}}, \bibinfo {author} {\bibfnamefont {A.}~\bibnamefont {Smirne}},
  \bibinfo {author} {\bibfnamefont {J.}~\bibnamefont {Lim}}, \bibinfo {author}
  {\bibfnamefont {S.~F.}\ \bibnamefont {Huelga}},\ and\ \bibinfo {author}
  {\bibfnamefont {M.~B.}\ \bibnamefont {Plenio}},\ }\href
  {https://doi.org/10.1103/PhysRevLett.123.090402} {\bibfield  {journal}
  {\bibinfo  {journal} {Phys. Rev. Lett.}\ }\textbf {\bibinfo {volume} {123}},\
  \bibinfo {pages} {090402} (\bibinfo {year} {2019})}\BibitemShut {NoStop}%
\bibitem [{\citenamefont {Prior}\ \emph {et~al.}(2010)\citenamefont {Prior},
  \citenamefont {Chin}, \citenamefont {Huelga},\ and\ \citenamefont
  {Plenio}}]{Prior2010}%
  \BibitemOpen
  \bibfield  {author} {\bibinfo {author} {\bibfnamefont {J.}~\bibnamefont
  {Prior}}, \bibinfo {author} {\bibfnamefont {A.~W.}\ \bibnamefont {Chin}},
  \bibinfo {author} {\bibfnamefont {S.~F.}\ \bibnamefont {Huelga}},\ and\
  \bibinfo {author} {\bibfnamefont {M.~B.}\ \bibnamefont {Plenio}},\ }\href
  {https://doi.org/10.1103/PhysRevLett.105.050404} {\bibfield  {journal}
  {\bibinfo  {journal} {Phys. Rev. Lett.}\ }\textbf {\bibinfo {volume} {105}},\
  \bibinfo {pages} {050404} (\bibinfo {year} {2010})}\BibitemShut {NoStop}%
\bibitem [{\citenamefont {Abramavicius}\ and\ \citenamefont
  {Abramavicius}(2014)}]{Abramavicius2014}%
  \BibitemOpen
  \bibfield  {author} {\bibinfo {author} {\bibfnamefont {V.}~\bibnamefont
  {Abramavicius}}\ and\ \bibinfo {author} {\bibfnamefont {D.}~\bibnamefont
  {Abramavicius}},\ }\href {https://doi.org/10.1063/1.4863968} {\bibfield
  {journal} {\bibinfo  {journal} {J. Chem. Phys}\ }\textbf {\bibinfo {volume}
  {140}},\ \bibinfo {pages} {065103} (\bibinfo {year} {2014})}\BibitemShut
  {NoStop}%
\bibitem [{\citenamefont {Appel}\ and\ \citenamefont {{Di
  Ventra}}(2009)}]{PhysRevB.80.212303}%
  \BibitemOpen
  \bibfield  {author} {\bibinfo {author} {\bibfnamefont {H.}~\bibnamefont
  {Appel}}\ and\ \bibinfo {author} {\bibfnamefont {M.}~\bibnamefont {{Di
  Ventra}}},\ }\href {https://doi.org/10.1103/PhysRevB.80.212303} {\bibfield
  {journal} {\bibinfo  {journal} {Phys. Rev. B}\ }\textbf {\bibinfo {volume}
  {80}},\ \bibinfo {pages} {212303} (\bibinfo {year} {2009})}\BibitemShut
  {NoStop}%
\bibitem [{\citenamefont {Biele}\ and\ \citenamefont
  {D'Agosta}(2012)}]{Biele_2012}%
  \BibitemOpen
  \bibfield  {author} {\bibinfo {author} {\bibfnamefont {R.}~\bibnamefont
  {Biele}}\ and\ \bibinfo {author} {\bibfnamefont {R.}~\bibnamefont
  {D'Agosta}},\ }\href {https://doi.org/10.1088/0953-8984/24/27/273201}
  {\bibfield  {journal} {\bibinfo  {journal} {J. Phys. Condens. Matter}\
  }\textbf {\bibinfo {volume} {24}},\ \bibinfo {pages} {273201} (\bibinfo
  {year} {2012})}\BibitemShut {NoStop}%
\bibitem [{\citenamefont {Di{\'{o}}si}\ and\ \citenamefont
  {Strunz}(1997)}]{Diosi1997}%
  \BibitemOpen
  \bibfield  {author} {\bibinfo {author} {\bibfnamefont {L.}~\bibnamefont
  {Di{\'{o}}si}}\ and\ \bibinfo {author} {\bibfnamefont {W.~T.}\ \bibnamefont
  {Strunz}},\ }\href {https://doi.org/10.1016/S0375-9601(97)00717-2} {\bibfield
   {journal} {\bibinfo  {journal} {Phys. Lett. Sect. A Gen. At. Solid State
  Phys.}\ }\textbf {\bibinfo {volume} {235}},\ \bibinfo {pages} {569} (\bibinfo
  {year} {1997})}\BibitemShut {NoStop}%
\bibitem [{\citenamefont {de~Vega}\ \emph {et~al.}(2005)\citenamefont
  {de~Vega}, \citenamefont {Alonso},\ and\ \citenamefont
  {Gaspard}}]{DeVega2005}%
  \BibitemOpen
  \bibfield  {author} {\bibinfo {author} {\bibfnamefont {I.}~\bibnamefont
  {de~Vega}}, \bibinfo {author} {\bibfnamefont {D.}~\bibnamefont {Alonso}},\
  and\ \bibinfo {author} {\bibfnamefont {P.}~\bibnamefont {Gaspard}},\ }\href
  {https://doi.org/10.1103/PhysRevA.71.023812} {\bibfield  {journal} {\bibinfo
  {journal} {Phys. Rev. A - At. Mol. Opt. Phys.}\ }\textbf {\bibinfo {volume}
  {71}},\ \bibinfo {pages} {023812} (\bibinfo {year} {2005})}\BibitemShut
  {NoStop}%
\bibitem [{\citenamefont {Link}\ and\ \citenamefont {Strunz}(2017)}]{Link2017}%
  \BibitemOpen
  \bibfield  {author} {\bibinfo {author} {\bibfnamefont {V.}~\bibnamefont
  {Link}}\ and\ \bibinfo {author} {\bibfnamefont {W.~T.}\ \bibnamefont
  {Strunz}},\ }\href {https://doi.org/10.1103/PhysRevLett.119.180401}
  {\bibfield  {journal} {\bibinfo  {journal} {Phys. Rev. Lett.}\ }\textbf
  {\bibinfo {volume} {119}},\ \bibinfo {pages} {180401} (\bibinfo {year}
  {2017})}\BibitemShut {NoStop}%
\bibitem [{\citenamefont {Hartmann}\ and\ \citenamefont
  {Strunz}(2017)}]{Hartmann2017}%
  \BibitemOpen
  \bibfield  {author} {\bibinfo {author} {\bibfnamefont {R.}~\bibnamefont
  {Hartmann}}\ and\ \bibinfo {author} {\bibfnamefont {W.~T.}\ \bibnamefont
  {Strunz}},\ }\href {https://doi.org/10.1021/acs.jctc.7b00751} {\bibfield
  {journal} {\bibinfo  {journal} {J. Chem. Theory Comput.}\ }\textbf {\bibinfo
  {volume} {13}},\ \bibinfo {pages} {5834} (\bibinfo {year}
  {2017})}\BibitemShut {NoStop}%
\bibitem [{\citenamefont {Reddy}\ and\ \citenamefont
  {Prasad}(2015)}]{Reddy2015}%
  \BibitemOpen
  \bibfield  {author} {\bibinfo {author} {\bibfnamefont {C.~S.}\ \bibnamefont
  {Reddy}}\ and\ \bibinfo {author} {\bibfnamefont {M.~D.}\ \bibnamefont
  {Prasad}},\ }\href {https://doi.org/10.1080/00268976.2015.1070928} {\bibfield
   {journal} {\bibinfo  {journal} {Mol. Phys.}\ }\textbf {\bibinfo {volume}
  {113}},\ \bibinfo {pages} {3023} (\bibinfo {year} {2015})}\BibitemShut
  {NoStop}%
\bibitem [{\citenamefont {Ritschel}\ \emph {et~al.}(2015)\citenamefont
  {Ritschel}, \citenamefont {Suess}, \citenamefont {M{\"{o}}bius},
  \citenamefont {Strunz},\ and\ \citenamefont {Eisfeld}}]{Ritschel2015}%
  \BibitemOpen
  \bibfield  {author} {\bibinfo {author} {\bibfnamefont {G.}~\bibnamefont
  {Ritschel}}, \bibinfo {author} {\bibfnamefont {D.}~\bibnamefont {Suess}},
  \bibinfo {author} {\bibfnamefont {S.}~\bibnamefont {M{\"{o}}bius}}, \bibinfo
  {author} {\bibfnamefont {W.~T.}\ \bibnamefont {Strunz}},\ and\ \bibinfo
  {author} {\bibfnamefont {A.}~\bibnamefont {Eisfeld}},\ }\href
  {https://doi.org/10.1063/1.4905327} {\bibfield  {journal} {\bibinfo
  {journal} {J. Chem. Phys.}\ }\textbf {\bibinfo {volume} {142}},\ \bibinfo
  {pages} {034115} (\bibinfo {year} {2015})}\BibitemShut {NoStop}%
\bibitem [{\citenamefont {Borrelli}\ and\ \citenamefont
  {Gelin}(2016)}]{Borrelli2016a}%
  \BibitemOpen
  \bibfield  {author} {\bibinfo {author} {\bibfnamefont {R.}~\bibnamefont
  {Borrelli}}\ and\ \bibinfo {author} {\bibfnamefont {M.~F.}\ \bibnamefont
  {Gelin}},\ }\href {https://doi.org/10.1063/1.4971211} {\bibfield  {journal}
  {\bibinfo  {journal} {J. Chem. Phys.}\ }\textbf {\bibinfo {volume} {145}},\
  \bibinfo {pages} {224101} (\bibinfo {year} {2016})}\BibitemShut {NoStop}%
\bibitem [{\citenamefont {Chen}\ and\ \citenamefont {Zhao}(2017)}]{Chen2017}%
  \BibitemOpen
  \bibfield  {author} {\bibinfo {author} {\bibfnamefont {L.}~\bibnamefont
  {Chen}}\ and\ \bibinfo {author} {\bibfnamefont {Y.}~\bibnamefont {Zhao}},\
  }\href {https://doi.org/10.1063/1.5000823} {\bibfield  {journal} {\bibinfo
  {journal} {J. Chem. Phys.}\ }\textbf {\bibinfo {volume} {147}},\ \bibinfo
  {pages} {214102} (\bibinfo {year} {2017})}\BibitemShut {NoStop}%
\bibitem [{\citenamefont {Martinazzo}\ \emph {et~al.}(2006)\citenamefont
  {Martinazzo}, \citenamefont {Nest}, \citenamefont {Saalfrank},\ and\
  \citenamefont {Tantardini}}]{Martinazzo2006a}%
  \BibitemOpen
  \bibfield  {author} {\bibinfo {author} {\bibfnamefont {R.}~\bibnamefont
  {Martinazzo}}, \bibinfo {author} {\bibfnamefont {M.}~\bibnamefont {Nest}},
  \bibinfo {author} {\bibfnamefont {P.}~\bibnamefont {Saalfrank}},\ and\
  \bibinfo {author} {\bibfnamefont {G.~F.}\ \bibnamefont {Tantardini}},\ }\href
  {https://doi.org/10.1063/1.2362821} {\bibfield  {journal} {\bibinfo
  {journal} {J. Chem. Phys.}\ }\textbf {\bibinfo {volume} {125}},\ \bibinfo
  {pages} {194102} (\bibinfo {year} {2006})}\BibitemShut {NoStop}%
\bibitem [{\citenamefont {Abramavicius}\ \emph {et~al.}(2018)\citenamefont
  {Abramavicius}, \citenamefont {Choro{\v{s}}ajev},\ and\ \citenamefont
  {Valkunas}}]{Abramavicius2018c}%
  \BibitemOpen
  \bibfield  {author} {\bibinfo {author} {\bibfnamefont {D.}~\bibnamefont
  {Abramavicius}}, \bibinfo {author} {\bibfnamefont {V.}~\bibnamefont
  {Choro{\v{s}}ajev}},\ and\ \bibinfo {author} {\bibfnamefont {L.}~\bibnamefont
  {Valkunas}},\ }\href {https://doi.org/10.1039/c8cp00682b} {\bibfield
  {journal} {\bibinfo  {journal} {Phys. Chem. Chem. Phys.}\ }\textbf {\bibinfo
  {volume} {20}},\ \bibinfo {pages} {21225} (\bibinfo {year}
  {2018})}\BibitemShut {NoStop}%
\bibitem [{\citenamefont {Chen}\ and\ \citenamefont
  {Sorbello}(1993)}]{Chen1993}%
  \BibitemOpen
  \bibfield  {author} {\bibinfo {author} {\bibfnamefont {Z.}~\bibnamefont
  {Chen}}\ and\ \bibinfo {author} {\bibfnamefont {R.~S.}\ \bibnamefont
  {Sorbello}},\ }\href {https://doi.org/10.1103/PhysRevB.47.13527} {\bibfield
  {journal} {\bibinfo  {journal} {Phys. Rev. B}\ }\textbf {\bibinfo {volume}
  {47}},\ \bibinfo {pages} {13527} (\bibinfo {year} {1993})}\BibitemShut
  {NoStop}%
\bibitem [{\citenamefont {Ichikawa}\ \emph {et~al.}(2007)\citenamefont
  {Ichikawa}, \citenamefont {Ichikawa}, \citenamefont {Yoshikawa},\ and\
  \citenamefont {Kimura}}]{Ichikawa2007}%
  \BibitemOpen
  \bibfield  {author} {\bibinfo {author} {\bibfnamefont {M.}~\bibnamefont
  {Ichikawa}}, \bibinfo {author} {\bibfnamefont {H.}~\bibnamefont {Ichikawa}},
  \bibinfo {author} {\bibfnamefont {K.}~\bibnamefont {Yoshikawa}},\ and\
  \bibinfo {author} {\bibfnamefont {Y.}~\bibnamefont {Kimura}},\ }\href
  {https://doi.org/10.1103/PhysRevLett.99.148104} {\bibfield  {journal}
  {\bibinfo  {journal} {Phys. Rev. Lett.}\ }\textbf {\bibinfo {volume} {99}},\
  \bibinfo {pages} {148104} (\bibinfo {year} {2007})}\BibitemShut {NoStop}%
\bibitem [{\citenamefont {Gulbinas}\ \emph {et~al.}(1996)\citenamefont
  {Gulbinas}, \citenamefont {Valkunas}, \citenamefont {Kuciauskas},
  \citenamefont {Katilius}, \citenamefont {Liuolia}, \citenamefont {Zhou},\
  and\ \citenamefont {Blankenship}}]{Gulbinas1996}%
  \BibitemOpen
  \bibfield  {author} {\bibinfo {author} {\bibfnamefont {V.}~\bibnamefont
  {Gulbinas}}, \bibinfo {author} {\bibfnamefont {L.}~\bibnamefont {Valkunas}},
  \bibinfo {author} {\bibfnamefont {D.}~\bibnamefont {Kuciauskas}}, \bibinfo
  {author} {\bibfnamefont {E.}~\bibnamefont {Katilius}}, \bibinfo {author}
  {\bibfnamefont {V.}~\bibnamefont {Liuolia}}, \bibinfo {author} {\bibfnamefont
  {W.}~\bibnamefont {Zhou}},\ and\ \bibinfo {author} {\bibfnamefont {R.~E.}\
  \bibnamefont {Blankenship}},\ }\href {https://doi.org/10.1021/jp961272k}
  {\bibfield  {journal} {\bibinfo  {journal} {J. Phys. Chem.}\ }\textbf
  {\bibinfo {volume} {100}},\ \bibinfo {pages} {17950} (\bibinfo {year}
  {1996})}\BibitemShut {NoStop}%
\bibitem [{\citenamefont {Valkunas}\ and\ \citenamefont
  {Gulbinas}(1997)}]{Valkunas1997}%
  \BibitemOpen
  \bibfield  {author} {\bibinfo {author} {\bibfnamefont {L.}~\bibnamefont
  {Valkunas}}\ and\ \bibinfo {author} {\bibfnamefont {V.}~\bibnamefont
  {Gulbinas}},\ }\href {https://doi.org/10.1111/j.1751-1097.1997.tb03199.x}
  {\bibfield  {journal} {\bibinfo  {journal} {Photochem. Photobiol.}\ }\textbf
  {\bibinfo {volume} {66}},\ \bibinfo {pages} {628} (\bibinfo {year}
  {1997})}\BibitemShut {NoStop}%
\bibitem [{\citenamefont {Gulbinas}\ \emph {et~al.}(2006)\citenamefont
  {Gulbinas}, \citenamefont {Karpicz}, \citenamefont {Garab},\ and\
  \citenamefont {Valkunas}}]{Gulbinas2006}%
  \BibitemOpen
  \bibfield  {author} {\bibinfo {author} {\bibfnamefont {V.}~\bibnamefont
  {Gulbinas}}, \bibinfo {author} {\bibfnamefont {R.}~\bibnamefont {Karpicz}},
  \bibinfo {author} {\bibfnamefont {G.}~\bibnamefont {Garab}},\ and\ \bibinfo
  {author} {\bibfnamefont {L.}~\bibnamefont {Valkunas}},\ }\href
  {https://doi.org/10.1021/bi060048a} {\bibfield  {journal} {\bibinfo
  {journal} {Biochemistry}\ }\textbf {\bibinfo {volume} {45}},\ \bibinfo
  {pages} {9559} (\bibinfo {year} {2006})}\BibitemShut {NoStop}%
\bibitem [{\citenamefont {Balevi{\v{c}}ius}\ \emph {et~al.}(2018)\citenamefont
  {Balevi{\v{c}}ius}, \citenamefont {Lincoln}, \citenamefont {Viola},
  \citenamefont {Cerullo}, \citenamefont {Hauer},\ and\ \citenamefont
  {Abramavicius}}]{Balevicius2018}%
  \BibitemOpen
  \bibfield  {author} {\bibinfo {author} {\bibfnamefont {V.}~\bibnamefont
  {Balevi{\v{c}}ius}}, \bibinfo {author} {\bibfnamefont {C.~N.}\ \bibnamefont
  {Lincoln}}, \bibinfo {author} {\bibfnamefont {D.}~\bibnamefont {Viola}},
  \bibinfo {author} {\bibfnamefont {G.}~\bibnamefont {Cerullo}}, \bibinfo
  {author} {\bibfnamefont {J.}~\bibnamefont {Hauer}},\ and\ \bibinfo {author}
  {\bibfnamefont {D.}~\bibnamefont {Abramavicius}},\ }\href
  {https://doi.org/10.1007/s11120-017-0423-6} {\bibfield  {journal} {\bibinfo
  {journal} {Photosynth. Res.}\ }\textbf {\bibinfo {volume} {135}},\ \bibinfo
  {pages} {55} (\bibinfo {year} {2018})}\BibitemShut {NoStop}%
\bibitem [{\citenamefont {{Balevi{\v{c}}ius Jr}}\ \emph
  {et~al.}(2019)\citenamefont {{Balevi{\v{c}}ius Jr}}, \citenamefont {Wei},
  \citenamefont {{Di Tommaso}}, \citenamefont {Abramavicius}, \citenamefont
  {Hauer}, \citenamefont {Pol{\'{i}}vka},\ and\ \citenamefont
  {Duffy}}]{Balevicius2019}%
  \BibitemOpen
  \bibfield  {author} {\bibinfo {author} {\bibfnamefont {V.}~\bibnamefont
  {{Balevi{\v{c}}ius Jr}}}, \bibinfo {author} {\bibfnamefont {T.}~\bibnamefont
  {Wei}}, \bibinfo {author} {\bibfnamefont {D.}~\bibnamefont {{Di Tommaso}}},
  \bibinfo {author} {\bibfnamefont {D.}~\bibnamefont {Abramavicius}}, \bibinfo
  {author} {\bibfnamefont {J.}~\bibnamefont {Hauer}}, \bibinfo {author}
  {\bibfnamefont {T.}~\bibnamefont {Pol{\'{i}}vka}},\ and\ \bibinfo {author}
  {\bibfnamefont {C.~D.~P.}\ \bibnamefont {Duffy}},\ }\href
  {https://doi.org/10.1039/C9SC00410F} {\bibfield  {journal} {\bibinfo
  {journal} {Chem. Sci.}\ }\textbf {\bibinfo {volume} {10}},\ \bibinfo {pages}
  {4792} (\bibinfo {year} {2019})}\BibitemShut {NoStop}%
\bibitem [{\citenamefont {Choi}\ \emph {et~al.}(2019)\citenamefont {Choi},
  \citenamefont {Zhou}, \citenamefont {Choi}, \citenamefont {Landig},
  \citenamefont {Ho}, \citenamefont {Isoya}, \citenamefont {Jelezko},
  \citenamefont {Onoda}, \citenamefont {Sumiya}, \citenamefont {Abanin},\ and\
  \citenamefont {Lukin}}]{Choi2019}%
  \BibitemOpen
  \bibfield  {author} {\bibinfo {author} {\bibfnamefont {J.}~\bibnamefont
  {Choi}}, \bibinfo {author} {\bibfnamefont {H.}~\bibnamefont {Zhou}}, \bibinfo
  {author} {\bibfnamefont {S.}~\bibnamefont {Choi}}, \bibinfo {author}
  {\bibfnamefont {R.}~\bibnamefont {Landig}}, \bibinfo {author} {\bibfnamefont
  {W.~W.}\ \bibnamefont {Ho}}, \bibinfo {author} {\bibfnamefont
  {J.}~\bibnamefont {Isoya}}, \bibinfo {author} {\bibfnamefont
  {F.}~\bibnamefont {Jelezko}}, \bibinfo {author} {\bibfnamefont
  {S.}~\bibnamefont {Onoda}}, \bibinfo {author} {\bibfnamefont
  {H.}~\bibnamefont {Sumiya}}, \bibinfo {author} {\bibfnamefont {D.~A.}\
  \bibnamefont {Abanin}},\ and\ \bibinfo {author} {\bibfnamefont {M.~D.}\
  \bibnamefont {Lukin}},\ }\href
  {https://doi.org/10.1103/PhysRevLett.122.043603} {\bibfield  {journal}
  {\bibinfo  {journal} {Phys. Rev. Lett.}\ }\textbf {\bibinfo {volume} {122}},\
  \bibinfo {pages} {043603} (\bibinfo {year} {2019})}\BibitemShut {NoStop}%
\bibitem [{\citenamefont {Scarani}\ \emph {et~al.}(2002)\citenamefont
  {Scarani}, \citenamefont {Ziman}, \citenamefont {{\v{S}}telmachovi{\v{c}}},
  \citenamefont {Gisin},\ and\ \citenamefont {Bu{\v{z}}ek}}]{Scarani2002}%
  \BibitemOpen
  \bibfield  {author} {\bibinfo {author} {\bibfnamefont {V.}~\bibnamefont
  {Scarani}}, \bibinfo {author} {\bibfnamefont {M.}~\bibnamefont {Ziman}},
  \bibinfo {author} {\bibfnamefont {P.}~\bibnamefont
  {{\v{S}}telmachovi{\v{c}}}}, \bibinfo {author} {\bibfnamefont
  {N.}~\bibnamefont {Gisin}},\ and\ \bibinfo {author} {\bibfnamefont
  {V.}~\bibnamefont {Bu{\v{z}}ek}},\ }\href
  {https://doi.org/10.1103/PhysRevLett.88.097905} {\bibfield  {journal}
  {\bibinfo  {journal} {Phys. Rev. Lett.}\ }\textbf {\bibinfo {volume} {88}},\
  \bibinfo {pages} {097905} (\bibinfo {year} {2002})}\BibitemShut {NoStop}%
\bibitem [{\citenamefont {Frenkel}\ and\ \citenamefont
  {Frenkel}(1936)}]{Frenkel1936}%
  \BibitemOpen
  \bibfield  {author} {\bibinfo {author} {\bibfnamefont {I.}~\bibnamefont
  {Frenkel}}\ and\ \bibinfo {author} {\bibfnamefont {J.}~\bibnamefont
  {Frenkel}},\ }\href@noop {} {\emph {\bibinfo {title} {{Wave Mechanics:
  Elementary Theory}}}}\ (\bibinfo  {publisher} {Oxford University Press},\
  \bibinfo {year} {1936})\BibitemShut {NoStop}%
\bibitem [{\citenamefont {Sun}\ \emph {et~al.}(2010)\citenamefont {Sun},
  \citenamefont {Luo},\ and\ \citenamefont {Zhao}}]{Sun2010a}%
  \BibitemOpen
  \bibfield  {author} {\bibinfo {author} {\bibfnamefont {J.}~\bibnamefont
  {Sun}}, \bibinfo {author} {\bibfnamefont {B.}~\bibnamefont {Luo}},\ and\
  \bibinfo {author} {\bibfnamefont {Y.}~\bibnamefont {Zhao}},\ }\href
  {https://doi.org/10.1103/PhysRevB.82.014305} {\bibfield  {journal} {\bibinfo
  {journal} {Phys. Rev. B - Condens. Matter Mater. Phys.}\ }\textbf {\bibinfo
  {volume} {82}},\ \bibinfo {pages} {014305} (\bibinfo {year}
  {2010})}\BibitemShut {NoStop}%
\bibitem [{\citenamefont {Zhang}\ \emph {et~al.}(1990)\citenamefont {Zhang},
  \citenamefont {Feng},\ and\ \citenamefont {Gilmore}}]{Zhang1990}%
  \BibitemOpen
  \bibfield  {author} {\bibinfo {author} {\bibfnamefont {W.-M.}\ \bibnamefont
  {Zhang}}, \bibinfo {author} {\bibfnamefont {D.~H.}\ \bibnamefont {Feng}},\
  and\ \bibinfo {author} {\bibfnamefont {R.}~\bibnamefont {Gilmore}},\ }\href
  {https://doi.org/10.1103/RevModPhys.62.867} {\bibfield  {journal} {\bibinfo
  {journal} {Rev. Mod. Phys}\ }\textbf {\bibinfo {volume} {62}},\ \bibinfo
  {pages} {867} (\bibinfo {year} {1990})}\BibitemShut {NoStop}%
\bibitem [{\citenamefont {Kais}\ and\ \citenamefont {Levine}(1990)}]{Kais1990}%
  \BibitemOpen
  \bibfield  {author} {\bibinfo {author} {\bibfnamefont {S.}~\bibnamefont
  {Kais}}\ and\ \bibinfo {author} {\bibfnamefont {R.~D.}\ \bibnamefont
  {Levine}},\ }\href {https://doi.org/10.1103/PhysRevA.41.2301} {\bibfield
  {journal} {\bibinfo  {journal} {Phys. Rev. A}\ }\textbf {\bibinfo {volume}
  {41}},\ \bibinfo {pages} {2301} (\bibinfo {year} {1990})}\BibitemShut
  {NoStop}%
\bibitem [{\citenamefont {Chen}\ \emph {et~al.}(2018)\citenamefont {Chen},
  \citenamefont {Gelin},\ and\ \citenamefont {Zhao}}]{Chen2018}%
  \BibitemOpen
  \bibfield  {author} {\bibinfo {author} {\bibfnamefont {L.}~\bibnamefont
  {Chen}}, \bibinfo {author} {\bibfnamefont {M.}~\bibnamefont {Gelin}},\ and\
  \bibinfo {author} {\bibfnamefont {Y.}~\bibnamefont {Zhao}},\ }\href
  {https://doi.org/10.1016/j.chemphys.2018.08.041} {\bibfield  {journal}
  {\bibinfo  {journal} {Chem. Phys.}\ }\textbf {\bibinfo {volume} {515}},\
  \bibinfo {pages} {108} (\bibinfo {year} {2018})}\BibitemShut {NoStop}%
\bibitem [{\citenamefont {Yan}\ \emph {et~al.}(2020)\citenamefont {Yan},
  \citenamefont {Chen}, \citenamefont {Luo},\ and\ \citenamefont
  {Zhao}}]{Yan2020}%
  \BibitemOpen
  \bibfield  {author} {\bibinfo {author} {\bibfnamefont {Y.}~\bibnamefont
  {Yan}}, \bibinfo {author} {\bibfnamefont {L.}~\bibnamefont {Chen}}, \bibinfo
  {author} {\bibfnamefont {J.~Y.}\ \bibnamefont {Luo}},\ and\ \bibinfo {author}
  {\bibfnamefont {Y.}~\bibnamefont {Zhao}},\ }\href
  {https://doi.org/10.1103/PhysRevA.102.023714} {\bibfield  {journal} {\bibinfo
   {journal} {Phys. Rev. A}\ }\textbf {\bibinfo {volume} {102}},\ \bibinfo
  {pages} {023714} (\bibinfo {year} {2020})}\BibitemShut {NoStop}%
\bibitem [{\citenamefont {Glauber}(1963)}]{Glauber1963}%
  \BibitemOpen
  \bibfield  {author} {\bibinfo {author} {\bibfnamefont {R.~J.}\ \bibnamefont
  {Glauber}},\ }\href {https://doi.org/10.1103/PhysRev.131.2766} {\bibfield
  {journal} {\bibinfo  {journal} {Phys. Rev.}\ }\textbf {\bibinfo {volume}
  {131}},\ \bibinfo {pages} {2766} (\bibinfo {year} {1963})}\BibitemShut
  {NoStop}%
\bibitem [{\citenamefont {Plenio}\ and\ \citenamefont
  {Knight}(1998)}]{Plenio1998}%
  \BibitemOpen
  \bibfield  {author} {\bibinfo {author} {\bibfnamefont {M.~B.}\ \bibnamefont
  {Plenio}}\ and\ \bibinfo {author} {\bibfnamefont {P.~L.}\ \bibnamefont
  {Knight}},\ }\href {https://doi.org/10.1103/revmodphys.70.101} {\bibfield
  {journal} {\bibinfo  {journal} {Rev. Mod. Phys.}\ }\textbf {\bibinfo {volume}
  {70}},\ \bibinfo {pages} {101} (\bibinfo {year} {1998})}\BibitemShut
  {NoStop}%
\bibitem [{\citenamefont {Luoma}\ \emph {et~al.}(2020)\citenamefont {Luoma},
  \citenamefont {Strunz},\ and\ \citenamefont {Piilo}}]{Luoma2020}%
  \BibitemOpen
  \bibfield  {author} {\bibinfo {author} {\bibfnamefont {K.}~\bibnamefont
  {Luoma}}, \bibinfo {author} {\bibfnamefont {W.~T.}\ \bibnamefont {Strunz}},\
  and\ \bibinfo {author} {\bibfnamefont {J.}~\bibnamefont {Piilo}},\ }\href
  {https://doi.org/10.1103/PhysRevLett.125.150403} {\bibfield  {journal}
  {\bibinfo  {journal} {Phys. Rev. Lett.}\ }\textbf {\bibinfo {volume} {125}},\
  \bibinfo {pages} {150403} (\bibinfo {year} {2020})}\BibitemShut {NoStop}%
\bibitem [{\citenamefont {Kampen}(2007)}]{Kampen2007}%
  \BibitemOpen
  \bibfield  {author} {\bibinfo {author} {\bibfnamefont {V.~N.}\ \bibnamefont
  {Kampen}},\ }\href {https://doi.org/10.1016/B978-0-444-52965-7.X5000-4}
  {\emph {\bibinfo {title} {Stochastic Processes in Physics and Chemistry}}}\
  (\bibinfo  {publisher} {Elsevier},\ \bibinfo {year} {2007})\BibitemShut
  {NoStop}%
\bibitem [{\citenamefont {Bertsekas}\ and\ \citenamefont
  {Tsitsiklis}(2008)}]{Bertsekas2008}%
  \BibitemOpen
  \bibfield  {author} {\bibinfo {author} {\bibfnamefont {D.}~\bibnamefont
  {Bertsekas}}\ and\ \bibinfo {author} {\bibfnamefont {J.}~\bibnamefont
  {Tsitsiklis}},\ }\href {https://doi.org/10.1887/0750305134/b558c5} {\emph
  {\bibinfo {title} {{Introduction to probability}}}}\ (\bibinfo  {publisher}
  {Athena Scientific},\ \bibinfo {year} {2008})\BibitemShut {NoStop}%
\bibitem [{\citenamefont {Kell}\ \emph {et~al.}(2013)\citenamefont {Kell},
  \citenamefont {Feng}, \citenamefont {Reppert},\ and\ \citenamefont
  {Jankowiak}}]{Kell2013a}%
  \BibitemOpen
  \bibfield  {author} {\bibinfo {author} {\bibfnamefont {A.}~\bibnamefont
  {Kell}}, \bibinfo {author} {\bibfnamefont {X.}~\bibnamefont {Feng}}, \bibinfo
  {author} {\bibfnamefont {M.}~\bibnamefont {Reppert}},\ and\ \bibinfo {author}
  {\bibfnamefont {R.}~\bibnamefont {Jankowiak}},\ }\href
  {https://doi.org/10.1021/jp405094p} {\bibfield  {journal} {\bibinfo
  {journal} {J. Phys. Chem. B}\ }\textbf {\bibinfo {volume} {117}},\ \bibinfo
  {pages} {7317} (\bibinfo {year} {2013})}\BibitemShut {NoStop}%
\bibitem [{\citenamefont {van Amerongen}\ \emph {et~al.}(2000)\citenamefont
  {van Amerongen}, \citenamefont {van Grondelle},\ and\ \citenamefont
  {Valkunas}}]{Amerongen2010}%
  \BibitemOpen
  \bibfield  {author} {\bibinfo {author} {\bibfnamefont {H.}~\bibnamefont {van
  Amerongen}}, \bibinfo {author} {\bibfnamefont {R.}~\bibnamefont {van
  Grondelle}},\ and\ \bibinfo {author} {\bibfnamefont {L.}~\bibnamefont
  {Valkunas}},\ }\href {https://doi.org/10.1142/9789812813664} {\emph {\bibinfo
  {title} {{Photosynthetic Excitons}}}}\ (\bibinfo  {publisher} {World
  Scientific},\ \bibinfo {year} {2000})\BibitemShut {NoStop}%
\bibitem [{\citenamefont {Choro{\v{s}}ajev}\ \emph {et~al.}(2016)\citenamefont
  {Choro{\v{s}}ajev}, \citenamefont {Rancova},\ and\ \citenamefont
  {Abramavicius}}]{Chorosajev2016c}%
  \BibitemOpen
  \bibfield  {author} {\bibinfo {author} {\bibfnamefont {V.}~\bibnamefont
  {Choro{\v{s}}ajev}}, \bibinfo {author} {\bibfnamefont {O.}~\bibnamefont
  {Rancova}},\ and\ \bibinfo {author} {\bibfnamefont {D.}~\bibnamefont
  {Abramavicius}},\ }\href {https://doi.org/10.1039/c5cp06871a} {\bibfield
  {journal} {\bibinfo  {journal} {Phys. Chem. Chem. Phys.}\ }\textbf {\bibinfo
  {volume} {18}},\ \bibinfo {pages} {7966} (\bibinfo {year}
  {2016})}\BibitemShut {NoStop}%
\bibitem [{\citenamefont {Moix}\ \emph {et~al.}(2012)\citenamefont {Moix},
  \citenamefont {Zhao},\ and\ \citenamefont {Cao}}]{Moix2012}%
  \BibitemOpen
  \bibfield  {author} {\bibinfo {author} {\bibfnamefont {J.~M.}\ \bibnamefont
  {Moix}}, \bibinfo {author} {\bibfnamefont {Y.}~\bibnamefont {Zhao}},\ and\
  \bibinfo {author} {\bibfnamefont {J.}~\bibnamefont {Cao}},\ }\href
  {https://doi.org/10.1103/PhysRevB.85.115412} {\bibfield  {journal} {\bibinfo
  {journal} {Phys. Rev. B}\ }\textbf {\bibinfo {volume} {85}},\ \bibinfo
  {pages} {115412} (\bibinfo {year} {2012})}\BibitemShut {NoStop}%
\bibitem [{\citenamefont {Subasi}\ \emph {et~al.}(2012)\citenamefont {Subasi},
  \citenamefont {Fleming}, \citenamefont {Taylor},\ and\ \citenamefont
  {Hu}}]{Subas2012}%
  \BibitemOpen
  \bibfield  {author} {\bibinfo {author} {\bibfnamefont {Y.}~\bibnamefont
  {Subasi}}, \bibinfo {author} {\bibfnamefont {C.~H.}\ \bibnamefont {Fleming}},
  \bibinfo {author} {\bibfnamefont {J.~M.}\ \bibnamefont {Taylor}},\ and\
  \bibinfo {author} {\bibfnamefont {B.~L.}\ \bibnamefont {Hu}},\ }\href
  {https://doi.org/10.1103/PhysRevE.86.061132} {\bibfield  {journal} {\bibinfo
  {journal} {Phys. Rev. E}\ }\textbf {\bibinfo {volume} {86}},\ \bibinfo
  {pages} {61132} (\bibinfo {year} {2012})}\BibitemShut {NoStop}%
\bibitem [{\citenamefont {Gelzinis}\ and\ \citenamefont
  {Valkunas}(2020)}]{Gelzinis2020}%
  \BibitemOpen
  \bibfield  {author} {\bibinfo {author} {\bibfnamefont {A.}~\bibnamefont
  {Gelzinis}}\ and\ \bibinfo {author} {\bibfnamefont {L.}~\bibnamefont
  {Valkunas}},\ }\href {https://doi.org/10.1063/1.5141519} {\bibfield
  {journal} {\bibinfo  {journal} {J. Chem. Phys.}\ }\textbf {\bibinfo {volume}
  {152}},\ \bibinfo {pages} {51103} (\bibinfo {year} {2020})}\BibitemShut
  {NoStop}%
\bibitem [{\citenamefont {Chin}\ \emph {et~al.}(2013)\citenamefont {Chin},
  \citenamefont {Prior}, \citenamefont {Rosenbach}, \citenamefont
  {Caycedo-Soler}, \citenamefont {Huelga},\ and\ \citenamefont
  {Plenio}}]{Chin2013a}%
  \BibitemOpen
  \bibfield  {author} {\bibinfo {author} {\bibfnamefont {A.~W.}\ \bibnamefont
  {Chin}}, \bibinfo {author} {\bibfnamefont {J.}~\bibnamefont {Prior}},
  \bibinfo {author} {\bibfnamefont {R.}~\bibnamefont {Rosenbach}}, \bibinfo
  {author} {\bibfnamefont {F.}~\bibnamefont {Caycedo-Soler}}, \bibinfo {author}
  {\bibfnamefont {S.~F.}\ \bibnamefont {Huelga}},\ and\ \bibinfo {author}
  {\bibfnamefont {M.~B.}\ \bibnamefont {Plenio}},\ }\href
  {https://doi.org/10.1038/nphys2515} {\bibfield  {journal} {\bibinfo
  {journal} {Nat. Phys.}\ }\textbf {\bibinfo {volume} {9}},\ \bibinfo {pages}
  {113} (\bibinfo {year} {2013})}\BibitemShut {NoStop}%
\bibitem [{\citenamefont {Wang}\ \emph {et~al.}(2016)\citenamefont {Wang},
  \citenamefont {Chen}, \citenamefont {Zhou},\ and\ \citenamefont
  {Zhao}}]{Wang2016a}%
  \BibitemOpen
  \bibfield  {author} {\bibinfo {author} {\bibfnamefont {L.}~\bibnamefont
  {Wang}}, \bibinfo {author} {\bibfnamefont {L.}~\bibnamefont {Chen}}, \bibinfo
  {author} {\bibfnamefont {N.}~\bibnamefont {Zhou}},\ and\ \bibinfo {author}
  {\bibfnamefont {Y.}~\bibnamefont {Zhao}},\ }\href
  {https://doi.org/10.1063/1.4939144} {\bibfield  {journal} {\bibinfo
  {journal} {J. Chem. Phys.}\ }\textbf {\bibinfo {volume} {144}},\ \bibinfo
  {pages} {024101} (\bibinfo {year} {2016})}\BibitemShut {NoStop}%
\bibitem [{\citenamefont {Baer}\ and\ \citenamefont
  {Kosloff}(1997)}]{Baer1997}%
  \BibitemOpen
  \bibfield  {author} {\bibinfo {author} {\bibfnamefont {R.}~\bibnamefont
  {Baer}}\ and\ \bibinfo {author} {\bibfnamefont {R.}~\bibnamefont {Kosloff}},\
  }\href {https://doi.org/10.1063/1.473950} {\bibfield  {journal} {\bibinfo
  {journal} {J. Chem. Phys.}\ }\textbf {\bibinfo {volume} {106}},\ \bibinfo
  {pages} {8862} (\bibinfo {year} {1997})}\BibitemShut {NoStop}%
\bibitem [{\citenamefont {Katz}\ \emph {et~al.}(2008)\citenamefont {Katz},
  \citenamefont {Gelman}, \citenamefont {Ratner},\ and\ \citenamefont
  {Kosloff}}]{Katz2008}%
  \BibitemOpen
  \bibfield  {author} {\bibinfo {author} {\bibfnamefont {G.}~\bibnamefont
  {Katz}}, \bibinfo {author} {\bibfnamefont {D.}~\bibnamefont {Gelman}},
  \bibinfo {author} {\bibfnamefont {M.~A.}\ \bibnamefont {Ratner}},\ and\
  \bibinfo {author} {\bibfnamefont {R.}~\bibnamefont {Kosloff}},\ }\href
  {https://doi.org/10.1063/1.2946703} {\bibfield  {journal} {\bibinfo
  {journal} {J. Chem. Phys.}\ }\textbf {\bibinfo {volume} {129}},\ \bibinfo
  {pages} {034108} (\bibinfo {year} {2008})}\BibitemShut {NoStop}%
\bibitem [{\citenamefont {Torrontegui}\ and\ \citenamefont
  {Kosloff}(2016)}]{Torrontegui2016a}%
  \BibitemOpen
  \bibfield  {author} {\bibinfo {author} {\bibfnamefont {E.}~\bibnamefont
  {Torrontegui}}\ and\ \bibinfo {author} {\bibfnamefont {R.}~\bibnamefont
  {Kosloff}},\ }\href {https://doi.org/10.1088/1367-2630/18/9/093001}
  {\bibfield  {journal} {\bibinfo  {journal} {New J. Phys.}\ }\textbf {\bibinfo
  {volume} {18}},\ \bibinfo {pages} {093001} (\bibinfo {year}
  {2016})}\BibitemShut {NoStop}%
\bibitem [{\citenamefont {Habecker}\ \emph {et~al.}(2019)\citenamefont
  {Habecker}, \citenamefont {R{\"{o}}hse},\ and\ \citenamefont
  {Kl{\"{u}}ner}}]{Habecker2019a}%
  \BibitemOpen
  \bibfield  {author} {\bibinfo {author} {\bibfnamefont {F.}~\bibnamefont
  {Habecker}}, \bibinfo {author} {\bibfnamefont {R.}~\bibnamefont
  {R{\"{o}}hse}},\ and\ \bibinfo {author} {\bibfnamefont {T.}~\bibnamefont
  {Kl{\"{u}}ner}},\ }\href {https://doi.org/10.1063/1.5119195} {\bibfield
  {journal} {\bibinfo  {journal} {J. Chem. Phys.}\ }\textbf {\bibinfo {volume}
  {151}},\ \bibinfo {pages} {134113} (\bibinfo {year} {2019})}\BibitemShut
  {NoStop}%
\bibitem [{\citenamefont {Zhou}\ \emph {et~al.}(2015)\citenamefont {Zhou},
  \citenamefont {Huang}, \citenamefont {Zhu}, \citenamefont {Chernyak},\ and\
  \citenamefont {Zhao}}]{Zhou2015b}%
  \BibitemOpen
  \bibfield  {author} {\bibinfo {author} {\bibfnamefont {N.}~\bibnamefont
  {Zhou}}, \bibinfo {author} {\bibfnamefont {Z.}~\bibnamefont {Huang}},
  \bibinfo {author} {\bibfnamefont {J.}~\bibnamefont {Zhu}}, \bibinfo {author}
  {\bibfnamefont {V.}~\bibnamefont {Chernyak}},\ and\ \bibinfo {author}
  {\bibfnamefont {Y.}~\bibnamefont {Zhao}},\ }\href
  {https://doi.org/10.1063/1.4923009} {\bibfield  {journal} {\bibinfo
  {journal} {J. Chem. Phys.}\ }\textbf {\bibinfo {volume} {143}},\ \bibinfo
  {pages} {014113} (\bibinfo {year} {2015})}\BibitemShut {NoStop}%
\bibitem [{\citenamefont {Zhou}\ \emph {et~al.}(2016)\citenamefont {Zhou},
  \citenamefont {Chen}, \citenamefont {Huang}, \citenamefont {Sun},
  \citenamefont {Tanimura},\ and\ \citenamefont {Zhao}}]{Zhou2016a}%
  \BibitemOpen
  \bibfield  {author} {\bibinfo {author} {\bibfnamefont {N.}~\bibnamefont
  {Zhou}}, \bibinfo {author} {\bibfnamefont {L.}~\bibnamefont {Chen}}, \bibinfo
  {author} {\bibfnamefont {Z.}~\bibnamefont {Huang}}, \bibinfo {author}
  {\bibfnamefont {K.}~\bibnamefont {Sun}}, \bibinfo {author} {\bibfnamefont
  {Y.}~\bibnamefont {Tanimura}},\ and\ \bibinfo {author} {\bibfnamefont
  {Y.}~\bibnamefont {Zhao}},\ }\href {https://doi.org/10.1021/acs.jpca.5b12483}
  {\bibfield  {journal} {\bibinfo  {journal} {J. Phys. Chem. A}\ }\textbf
  {\bibinfo {volume} {120}},\ \bibinfo {pages} {1562} (\bibinfo {year}
  {2016})},\ \Eprint {https://arxiv.org/abs/1908.09243} {arXiv:1908.09243}
  \BibitemShut {NoStop}%
\bibitem [{\citenamefont {Jaku{\v{c}}ionis}\ \emph {et~al.}(2020)\citenamefont
  {Jaku{\v{c}}ionis}, \citenamefont {Mancal},\ and\ \citenamefont
  {Abramavi{\v{c}}ius}}]{Jakucionis2020}%
  \BibitemOpen
  \bibfield  {author} {\bibinfo {author} {\bibfnamefont {M.}~\bibnamefont
  {Jaku{\v{c}}ionis}}, \bibinfo {author} {\bibfnamefont {T.}~\bibnamefont
  {Mancal}},\ and\ \bibinfo {author} {\bibfnamefont {D.}~\bibnamefont
  {Abramavi{\v{c}}ius}},\ }\href {https://doi.org/10.1039/d0cp01092h}
  {\bibfield  {journal} {\bibinfo  {journal} {Phys. Chem. Chem. Phys.}\
  }\textbf {\bibinfo {volume} {22}},\ \bibinfo {pages} {8952} (\bibinfo {year}
  {2020})}\BibitemShut {NoStop}%
\end{thebibliography}%

\end{document}